\documentclass[a4paper,11pt]{article}
\pdfoutput=1
\usepackage{jcappub}
\usepackage{epsfig}
\usepackage{caption}
\usepackage{subcaption}
\usepackage{color,soul}
\usepackage{float}
\usepackage{amsfonts}
\usepackage{url}
\usepackage{slashed}
\usepackage{accents}
\usepackage{lipsum}
\usepackage{mathtools}
\usepackage{physics}

\usepackage{orcidlink}

\allowdisplaybreaks

\title{\boldmath Superradiant and dynamical spin-down of neutron stars with gravitational wave implications}

\author[a,b,c]{Indra Kumar Banerjee$^{\orcidlink{https://orcid.org/0000-0003-3900-735X}}$,}
\author[c]{Sandeep Chatterjee$^{\orcidlink{https://orcid.org/0000-0003-3914-782X}}$,}
\author[c]{Biswarup Das$^{\orcidlink{https://orcid.org/0009-0000-8486-8795}}$,}
\author[c]{Ujjal Kumar Dey$^{\orcidlink{https://orcid.org/0000-0002-9620-7561}}$}
\affiliation[a]{Institute of Mathematical Sciences, HBNI,\\ CIT Campus, Chennai 600113, India}
\affiliation[b]{Homi Bhabha National Institute,\\ Training School Complex, Anushakti Nagar, Mumbai 400085, India}
\affiliation[c]{Department of Physical Sciences, Indian Institute of Science Education and Research Berhampur,\\Berhampur, Odisha 760003, India}

\emailAdd{indrakumarb@imsc.res.in}
\emailAdd{sandeep@iiserbpr.ac.in}
\emailAdd{biswarupd21@iiserbpr.ac.in}
\emailAdd{ujjal@iiserbpr.ac.in}

\abstract{Neutron stars such as pulsars and magnetars lose angular momentum primarily through electromagnetic dipole radiation, gravitational waves, $r$-mode oscillation. However, spin variations, particularly enhanced spin-down rates, indicate additional spin-down mechanisms. We propose superradiant spin-down as a potential faster spin-down process. By modelling the interplay between conventional and superradiant spin-down channels, we evaluate their impact on neutron star rotational evolution. We also discuss gravitational-wave emission produced by quadrupole deformation, $r$-mode oscillations, and axion-induced bosonic clouds around an isolated neutron star, highlighting their potential as distinct multi-messenger probes in upcoming detectors.}

\begin{document}
\maketitle
\flushbottom


\section{Introduction}
\label{sec:intro}
The observable characteristics of neutron stars are governed primarily by their spin period and magnetic field. Understanding how the spin evolves is key to interpreting existing observations and uncovering potential new astrophysical phenomena~\cite{Abolmasov:2024nwa}. Rotational evolution of a neutron star is governed by various spin-down mechanisms that collectively determine how its spin frequency changes over time. A key diagnostic of this evolution is the braking index $n$ which can be expressed as $n=\Omega\ddot{\Omega}/{\dot{\Omega}}^2 $~\cite{Johnston:1999ka}, where
\( \dot{\Omega} \) and \( \ddot{\Omega} \) correspond to the first and second time derivatives of rotational spin angular frequency of the star, capturing the instantaneous spin-down rate.
Observations of pulsars and magnetars show a broad range of measured braking indices~\cite{Lyne:2014qqa}. 
Magnetars, with their extremely strong dipolar magnetic fields \( B \approx 10^{13} - 10^{15} \, \text{G} \), predominantly lose rotational energy through electromagnetic radiation, for which the theoretically expected braking index is 
\( n = 3 \)~\cite{Hamil:2015hqa}.
In scenarios where an alternative spin-down mechanism, such as gravitational wave (GW) emission, dominates over magnetic dipole radiation -- commonly referred to as the gravitar scenario~\cite{Palomba:2005na}, the braking index can reach values as high as \( n = 5\)~\cite{Hamil:2015hqa}. In such GW-dominated cases, the equatorial ellipticity \( \epsilon \) plays a crucial role, as larger values of \( \epsilon \) lead to a significantly faster spin-down rate. 
Another standard spin-down mechanism $r$-mode instabilities are non-radial oscillations in rotating fluids, which are driven by the Coriolis force. These instabilities highly depend on the $r$-mode oscillation amplitude $\alpha$ (remains constant during saturated phase), which can cause significant additional spin-down, leading to a braking index of \(n=7\)~\cite{Andersson:2000mf}.
However, interplay of multiple competing mechanisms, require a broader perspective beyond the conventional braking index.
In this work, we explore additional non-standard spin-down mechanisms, superradiance~\cite{Day:2019bbh} as a potential candidate contributing to the spin-down of isolated neutron stars.

\begin{figure}
	\centering
	\includegraphics[scale=0.5]{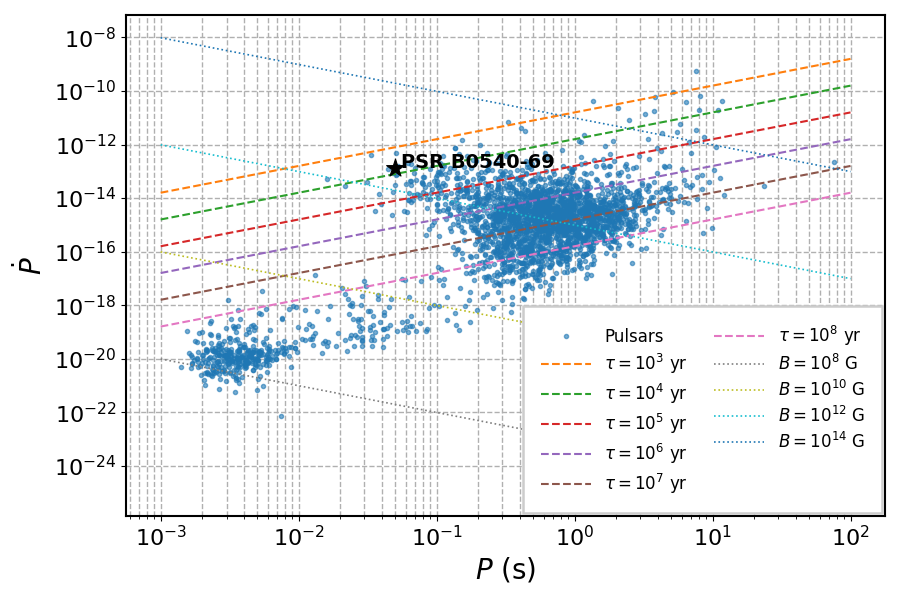}
	\caption{Pulsar spin parameters $P$-$\dot{P}$ distribution from the ATNF catalogue~\cite{Manchester:2004bp}.}
	\label{fig:pp}
\end{figure}

Superradiance is a phenomenon in stars, where incoming waves are amplified by transferring angular momentum from the star's rotation. Moreover, in the presence of ultralight bosonic fields, superradiance can induce an instability that extracts angular momentum from the neutron star, thereby contributing to its spin-down during the instability phase. 
Similar to black holes, neutron stars generate a Schwarzschild-like gravitational potential that can support axion bound states around the star. While black holes lose energy through their ergoregion, neutron stars-owing to their high bulk conductivity~\cite{Potekhin:2001id} -- act as dissipative media, enabling superradiant extraction of angular momentum as long as the stellar rotational frequency \( \Omega \) exceeds the axion bound-state frequency \( \omega \)~\cite{Brito:2014wla, Day:2019bbh}.
In this work, we consider superradiant instability as a potential spin-down mechanism that may account for deviations from the standard values of the braking index and also which can explain faster spin-down rate observed in neutron stars. Spin-down via this mechanism is highly governed by parameters such as the magnetic field strength, axion mass, axion-photon coupling, stellar rotation rate, and the bulk conductivity of the star -- all of which collectively determine the characteristic timescale of the superradiant instability. Moreover, the resulting axionic clouds can act as sources of continuous gravitational wave emission through the process of axion annihilation and level transitions. Axion annhilation involves pairs of axions converting directly into gravitons, while level transitions occur when axions move between bound states. We focus on GW from axion annihilation since it generates significantly stronger signals compared to level transitions.
Key observable quantities of pulsar and magnetar, such as spin period $P$ and its derivative $\dot{P}$ are available in major catalogues such as ATNF pulsar catalogue and McGill magnetar catalogue~\cite{Manchester:2004bp,Olausen:2013bpa}. In Fig.~\ref{fig:pp}, $P$-$\dot{P}$ distribution from ATNF pulsar catalogue~\cite{Manchester:2004bp} is shown, with each data point representing a detected pulsar. Considering dominant spin-down via electromagnetic radiation, the reference lines denote constant magnetic field strengths and characteristic lifetimes. The black star marker corresponds to the pulsar PSR B0540-69, a pulsar known for its anti-glitch events.

The article is organized as follows. In Section \ref{sec:spinDwnMech}, we examine the principal spin-down mechanisms and superradiance, and assess their combined impact on neutron star spin-period evolution under various astrophysical conditions. Section~\ref{sec:gwSpec} analyzes the characteristic gravitational wave emission produced by quadrupolar deformation and fluid oscillation, including continuous waves from superradiance induced axion clouds. These provide strain estimates and detectability prospects for present and next-generation detectors~\cite{Hild:2011np, Borhanian:2022czq}. Finally we summarize our results and conclude in Section~\ref{sec:concl}.

\section{Mechanisms of spin-down}
\label{sec:spinDwnMech}
Neutron star loses its rotational energy through several standard dissipative mechanisms. These processes collectively shape the long-time rotational behaviour of stars. In addition to these conventional channels, spin-down via superradiance has emerged as a compelling non-standard mechanism capable of explaining anomalous phenomena, such as enhanced spin-down rate and corresponding braking index values~\cite{Tuo:2024pvf}. Together, these pathways shape the spin evolution of neutron stars and offer clues about their internal physics. 

\subsection{Electromagnetic radiation}
\label{sbsec:elecemsn}
A neutron star hosts an intense magnetic field that interacts with its surroundings. For pulsars and magnetars, electromagnetic radiation serves as the dominant mechanism driving their rotational spin-down. Considering a neutron star of radius $R$ with a dipolar magnetic field of polar strength $B$, whose magnetic axis is inclined at an angle $\theta$ to the rotation axis. The misalignment causes the magnetic dipole moment $m$ to vary with time~\cite{Abolmasov:2024nwa}. 
In realistic situations the neutron star magnetosphere is expected to be filled with plasma, which modifies the spin-down torque compared to the ideal vacuum dipole case and weakens its dependence on the magnetic inclination angle. Force-free magnetosphere simulations show that the electromagnetic energy loss remains non-zero even for an aligned rotator and is approximately proportional to $(1+\sin^2\theta)$. Adopting this more realistic description of the magnetospheric torque, the rate of rotational energy loss due to electromagnetic radiation can be written as~\cite{Spitkovsky:2006np}
\begin{gather}
	\dot{E}  
	= -\frac{\mu^{2} \Omega^4 (1+\sin^2\theta)}{c^3},\\
	\Rightarrow ~~\dot{\Omega} \propto -\Omega^3 ,
\label{2.1}
\end{gather}
where magnetic dipole moment $\mu = BR^{3}/2$.
If the star loses rotational energy primarily through electromagnetic dipole radiation and the magnetic field strength and inclination angle remain constant, the resulting spin-down follows a braking index of $n=3$~\cite{Hamil:2015hqa}.
\subsection{Gravitational wave emission}
\label{sbsec:gwemsn}
A perfectly spherical isolated neutron star would not emit any gravitational waves. However, if the star has deformations, such as a slight bulge (often referred to as a ``mountain"), uneven mass distribution, or precession, these can generate time-varying quadrupole moments, leading to the production of gravitational wave signals~\cite{Huang:2022vnz}. 
These deformations can arise from various factors, including magnetic fields~\cite{Mastrano:2011tf}, starquakes~\cite{Giliberti:2022uwc}, or accretion~\cite{Piro:2011ed} processes. For magnetically induced deformation, the density distribution of the star is affected by the asymmetric perturbation of magnetic stress. As the neutron star emits gravitational waves, it loses energy, which is drawn directly from the star's rotational energy, resulting in a gradual spin-down.
The rate at which energy is lost due to gravitational waves is primarily related to the star's equatorial ellipticity \(\epsilon\), moment of inertia \(I\), and its rotation rate \(\Omega\), which can be defined by the following equation~\cite{Lasky:2015olc},
\begin{gather}
	\Dot{E}_{\mathrm{GW}(m)} = -\frac{32}{5} \frac{GI^2}{c^5} \epsilon^2 \Omega^6,
	\label{2.3}
	\\
	\Rightarrow ~~\dot{\Omega} \propto -\Omega^{5}.
\end{gather}
Equatorial ellipticity $\epsilon$ can be affected by magnetic fields, starquakes or accretion. For highly magnetized stars magnetically induced ellipticity becomes more effective, which can be defined as $\epsilon_{\mathrm{mag}}$~\cite{Huang:2022vnz,Mastrano:2011tf}:
\begin{equation}
\epsilon_{\mathrm{mag}} = \pi I^{-1} \int_v \delta\rho_{\mathrm{mag}}(r,\theta') r^4 \sin\theta' (1 - 3\cos^2\theta') dr d\theta',
\end{equation}
where $\delta\rho_{\mathrm{mag}}(r,\theta')$ denotes the profile of density perturbation due to magnetic stress. 
Assuming a typical value $\epsilon \approx 10^{-6}$ ensures that the gravitational-wave-induced spin-down is comparable in magnitude to the other spin-down mechanisms, allowing us to investigate the interplay between the different spin-down processes.
In a simplified treatment, gravitational-wave dominated spin-down corresponds to a braking index \(n = 5\)~\cite{Hamil:2015hqa}.
%

%
\subsection{$r$-mode oscillation}
\label{sbsec:rmode}
$r$-modes are a type of non-radial oscillation that occur in rotating fluids, that are driven primarily by the Coriolis force, which acts as the restoring force due to the star's rotation. Instability of $r$-modes also can be driven by gravitational radiation~\cite{Lockitch:1998nq}. Eulerian velocity perturbation of the $r$-mode is defined as $\delta v=\alpha \Omega R(r/R)^{l} \vec{Y}^{B}_{lm}e^{i\omega t}$, where $\alpha$ is amplitude of the mode~\cite{Lockitch:1998nq} and the magnetic-type vector spherical harmonic of order $l$, $m$ can be written as $\vec{Y}^{B}_{lm}=[l(l+1)]^{-1/2}r\vec{\nabla}\times(r\vec{\nabla} Y_{lm})$~\cite{Owen:1998xg}. Assuming that the angular momentum ($J$) of the star is only dissipated via gravitational waves following current multipole formula ($l=m=2$), we get $\dot{J}_{\mathrm{GW}}=3\Omega\alpha^2\Tilde{J}MR^{2}/ \tau_{r} $~\cite{Andersson:2000mf}. Here $\Tilde{J}$ corresponds to the dimensionless constant that encodes how stellar mass is distributed and how it participates in $r$-mode motion, where $\tau_{r}$ represents the instability timescale of $r$-mode oscillations.
Explicitly, $r$-mode oscillation can be described in two stages -- early phase and late phase. Early phase is the growing phase of $r$-mode amplitude, where it increases due to instability $\dot{\alpha}\neq0$. Spin-down rate during the early phase of $r$-mode oscillation follows~\cite{Andersson:2000mf},
\begin{equation}\dot\Omega=-\frac{3\alpha^2\Tilde{J}\Omega}{(\Tilde{I}+\frac{3}{2}\alpha^{2}\Tilde{J})\tau_{\mathrm{diss}}},
\end{equation}
where dissipative timescale $\tau_{\mathrm{diss}}$ can be defined as,
\begin{equation}\frac{1}{\tau_{\mathrm{diss}}}=\frac{T^{-2}_{9}}{\tau_{s}}+\frac{T^{6}_{9}}{\tau_{b}}\Omega^{2}.
\end{equation}
Here $\tau_{s}$ and $\tau_{b}$ denote the damping timescales associated with shear viscosity and bulk viscosity respectively. The quantity $T_{9}$ represents the temperature scaled as $(T/10^{9}\mathrm{K})$. 

For late phase as $r$-mode stabilizes, $\alpha$ stays constant throughout the saturated phase, considering $\dot{\alpha}=0$. The energy loss due to gravitational waves from $r$-mode oscillations can be approximated as~\cite{Andersson:2000mf},
\begin{gather}
	\Dot{E}_{\mathrm{GW}(r)} \approx -\frac{G M R^4 \Tilde{I}^2 \Tilde{J}^2 \alpha^2 \Omega^8}{c^7}, \label{2.8}
	\\
	\Rightarrow ~~\dot{\Omega} \propto -\Omega^{7}.
\end{gather}
\begin{figure}[t]
	\centering
	\includegraphics[scale=0.75]{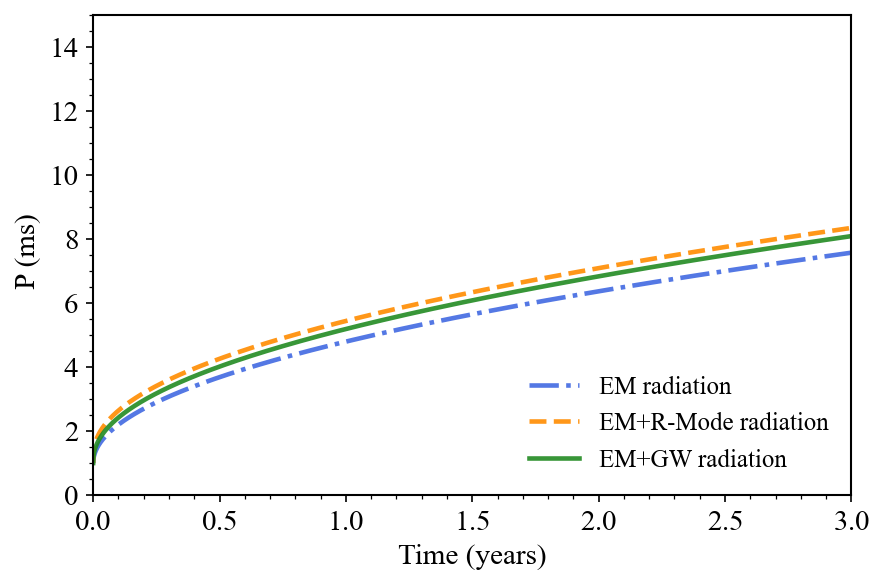}
	\caption{Spin period evolution of an isolated milisecond neutron star due to different spindown mechanisms.}
	\label{fig:pp1}
\end{figure}
Fig.~\ref{fig:pp1} shows the spin-period evolution of an isolated star due to different mechanisms, assuming magnetic field $B=10^{13}$G, $r$-mode amplitude $\alpha=10^{-4}$ and ellipticity $\epsilon\approx10^{-6}$~\cite{Andersson:2000mf}. Here dimensionless parameters $\Tilde{J} = 0.01635$ and $\Tilde{I} = I/(MR^2) = 0.261$ are obtained using a $\Gamma = 2$ polytropic model with a stellar radius of $R = 10\,\mathrm{km}$~\cite{Friedman:1997uh}.  
Under these assumptions, the braking index corresponds to $n=7$.


\subsection{Superradiance}
\label{sbsec:suprad}
In astrophysical context superradiance refers to the extraction of energy from a compact object through the amplification of surrounding fields. Although several conditions must be satisfied for this process to operate, it becomes particularly efficient in the presence of bosonic fields with non-zero mass. For our discussion we consider this bosonic field to be axionic field. For a rotating compact object superradiance extracts angular momentum, triggering the amplification of axionic modes and this transfer of rotational energy to the field causes the object to spin-down over time.
The basic process is based on two features, i.e., a rotating background which provides a source for angular momentum and a dissipative mechanism due to which the wave can perform some work on the rotation and gain energy. 
In the more traditional picture for Kerr black holes (BH) the ergo-region provides the dissipative mechanism and discussed in great detail in the literature. In the case of neutron stars, there is no ergo-region but one can rely on the magnetosphere of the neutron star for this purpose \cite{Day:2019bbh}. The magnetosphere, which is filled with charged plasma, co-rotates (almost) with the neutron star and it possesses an effective bulk conductivity ($\sigma_{M}$). However, unlike the Kerr BH, here only electromagnetic waves experience direct dissipation. When an electromagnetic fluctuation propagates through the co-rotating plasma, the electric field associated with the fluctuation gives rise to the current in the plasma which can dissipate energy due to the bulk conductivity. As the plasma is rotating, this energy dissipation is not symmetric and depending on the frequency ($\omega$) and the azimuthal mode ($m^{\prime}$) of the wave, it can either lose or gain energy. In the rotating frame of the plasma the frequency of the wave is Doppler shifted and can be expressed as $\omega^{\prime} = \omega - m^{\prime}\Omega$, where $\Omega$ is the angular frequency of the plasma. Now for the $\omega^{\prime} > 0$ the plasma perceives that the wave has a positive energy and damps the wave. However for $\omega^{\prime} < 0$, the plasma experiences a wave with `negative' energy and hence for this case when the dissipation occurs a negative energy is removed from the system which leads to the amplification of the wave. In general,  if an electromagnetic wave gets amplified, then it may just radiate away the energy. However, if there exists a coupling between axion and photons $g_{A\gamma\gamma}$ then some of this energy is transferred into the axions and which leads to the amplification of the axion waves. This physically leads to the increase in the occupation number of the axions. Neutron stars generate an external Schwarzschild gravitational potential that allows for hydrogen-like bound state solutions of this massive axion field $A$. The axionic field around a rotating neutron star can form gravitationally bound states analogous to the electronic bound states of the hydrogen atom. These states are described by wave functions characterized by the quantum numbers $(n^{\prime},l^{\prime},m^{\prime})$, where $n^{\prime}, l^{\prime},$ and $m^{\prime}$ are the principal, orbital, and azimuthal quantum numbers, respectively.
The strength of the gravitational interaction between the neutron star and the bosonic particle is characterized by the dimensionless parameter $\alpha_{0}$, which plays the role of the gravitational analogue of the electromagnetic fine-structure constant. This is defined as~\cite{Hessels:2006ze},
\begin{align}
	\alpha_{0} = \frac{G M_{\mathrm{NS}} m_{A}}{\hbar c^{3}}
	\approx 0.00105
	\left(
	\frac{m_{A}}{10^{-13}\,\mathrm{eV}}
	\right),
\end{align}
where $G$ is the Newton's gravitational constant, $M_{\mathrm{NS}}$ is the neutron star mass, $m_A$ is the boson mass, $\hbar$ is the reduced Planck constant.
The above-mentioned bound state manifests itself as a cloud of axions with rest energy $m_{A}$. For a neutron star of mass $M_\mathrm{{NS}}$ this occurs when the superradiant condition $\Omega > m^{\prime} \omega$ is satisfied, and during the superradiance instability phase the neutron star transfers its mass and spin to the scalar field. Neglecting accretion for simplicity, the evolution of the system is governed by coupled dynamics of the neutron star and the axion cloud~\cite{Brito:2017zvb}. The interaction between the axion and electromagnetic fields is described by the Lagrangian 
\begin{align}
	\mathcal{L} \supset 
	\frac{1}{2} \partial_\mu A \, \partial^\mu A
	- \frac{1}{2}m_A^2 A^2
	- \frac{1}{4} F_{\mu\nu}\tilde{F}^{\mu\nu} - \frac{g_{A\gamma\gamma}}{4} \phi F_{\mu\nu} \tilde{F}^{\mu\nu}
	- A_\mu j^\mu,
\end{align}
where $g_{A\gamma\gamma}$ is the axion-photon coupling, and $j^\mu$ is the current in the surrounding medium, defined as,
\begin{align}
	j^\mu = \sigma F^{\mu\nu}u_\nu + \rho \, u^\mu, 
	\quad u^\mu u_\mu = 1,
\end{align}
where $\rho$ is the electric charge density, $\sigma$ the conductivity, and $u^\mu$ the four-velocity of the medium. The presence of axion-photon coupling can effectively characterized by a perturbation in the axion bound state frequency $\omega_{l^{\prime}m^{\prime}n^{\prime}}$~\cite{Day:2019bbh}. 
The assumption of a uniform magnetic field, $\mathbf{B}=B\hat{\mathbf{z}}$, is adopted as a simplifying approximation that captures the leading-order physics of axion--photon mixing while preserving axial symmetry. This is reasonable because the dominant superradiant mode ($l^{\prime}=m^{\prime}=1$) is localized near the stellar equator, where the magnetic field of an aligned dipole is approximately directed along the $z$-axis. For simplicity, considering the assumption of magnetic field $\vec{B}$ along $z$-axis, the imaginary part of the frequency, which represents the instability, is given by~\cite{Day:2019bbh, Cardoso:2017kgn},
\begin{align}
	\mathrm{Im}[\omega_{l^{\prime}m^{\prime}n^{\prime}}]  \simeq \frac{\pi g^2_{A\gamma\gamma} B^2}{32\omega^3} & \sigma_M \alpha_{0}^{2l^{\prime}+5}F_{l^{\prime}n^{\prime}}  \left((m^{\prime}\Omega-\omega)-\omega S[l^{\prime},l^{\prime}+1,m^{\prime}]^2\right) \nonumber \\ &\times \left((R_{LC}\omega)^{2l^{\prime}+3}-(R\omega)^{2l^{\prime}+3}\right),
\end{align}
where 
\begin{equation}
	F_{l^{\prime}n^{\prime}} = \frac{2\Gamma(l^{\prime}+2)^2\Gamma(2l^{\prime}+n^{\prime}+2)}{(2l^{\prime}+3)n^{\prime}!(l^{\prime}+n^{\prime}+1)\Gamma(l^{\prime}+\frac{3}{2})^2\Gamma(2l^{\prime}+2)^2},
\end{equation}
and $S$ is a Wigner $3j$ symbol encoding angular momentum coupling. For the transitions $l^{\prime}=1$ to $l^{\prime}=2$, it can be simplified as~\cite{Day:2019bbh},
\begin{align}
	S[l^{\prime},l^{\prime}+1,m^{\prime}=l^{\prime}]^2= \frac{1}{(3+2l^{\prime})}.
\end{align}
\begin{figure}[t]
	\centering
	\includegraphics[scale=0.45]{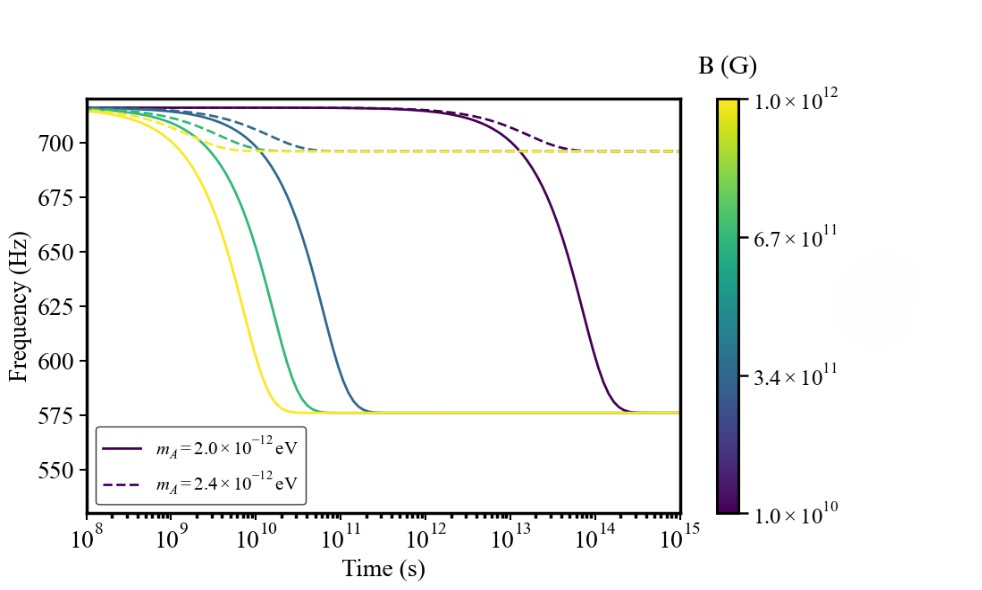}
	\caption{Spin-down evolution during superradiant instability for a canonical neutron star, assuming $g_{A\gamma\gamma}=10^{-12}\,\mathrm{GeV}^{-1}$.}
	\label{fig:pp4}
\end{figure}
Here $R_{LC}$ represent the light-cylinder radius. The basic conditions required for superradiance in neutron stars are given by $g_{A\gamma\gamma} B \ll \omega$ and $\Omega \gg \omega$. The superradiant instability timescale can be defined as the inverse of the imaginary part of the frequency,
\begin{equation}
	\tau_{\mathrm{inst}}
	=
	\frac{1}{\mathrm{Im}[\omega_{l^{\prime}m^{\prime}n^{\prime}}]}.
\end{equation}
To estimate the maximum possible spin-down effect arising solely from superradiance, we equate the instability timescale $\tau_{\mathrm{inst}}
=1/\mathrm{Im}[\omega_{l^{\prime}m^{\prime}n^{\prime}}]$ with the characteristic age of the neutron star $ 	\tau_{\mathrm{age}}$, which is defined as $\tau_{\mathrm{age}}=-\Omega/\dot{\Omega}$.
This gives the spin-down rate due to superradiance as 
\begin{align}
	\dot{\Omega} = - \Omega\,\mathrm{Im}[\omega_{l^{\prime}m^{\prime}n^{\prime}}]  \simeq \frac{\pi g^2_{A\gamma\gamma} \Omega B^2}{32\omega^3} & \sigma_M \alpha_{0}^{2l^{\prime}+5}F_{l^{\prime}n^{\prime}}  \left((m^{\prime}\Omega-\omega)-\omega S[l^{\prime},l^{\prime}+1,m^{\prime}]^2\right) \nonumber \\ &\times \left((R_{LC}\omega)^{2l^{\prime}+3}-(R\omega)^{2l^{\prime}+3}\right),
	\label{2.17}
\end{align}
 Equation~\eqref{2.17} can be recast into a spin-down rate expression of the form
\begin{align}
	\dot{\Omega}\propto \Omega^{-4}\left(\Omega - \frac{6\omega}{5}\right)\omega^{2}B^{2}m_{A}^{7},
	\label{2.18}
\end{align}
which captures the characteristic dependence of the instability-driven spin-down on the bound state boson's frequency $\omega$. Using the standard relation between the spin frequency $f$ and the angular frequency, $\Omega = 2\pi f$, Eq.~\eqref{2.18} can be rewritten in a form suitable for astrophysical estimates as
\begin{align}
	\dot{f}
	&\approx
	-10^{-6}\,\mathrm{Hz\,s^{-1}}
	\left(\frac{\,f-\dfrac{6\omega}{5}}{1\,\mathrm{Hz}}\right)
	\left(\frac{f}{1\,\mathrm{Hz}}\right)^{-4}
	\left(\frac{\omega}{1\,\mathrm{Hz}}\right)^{2}
	\left(\frac{B}{10^{9}\,\mathrm{G}}\right)^{2}
	\left(\frac{m_{A}}{10^{-11}\,\mathrm{eV}}\right)^{7}.
	\label{eq:fdot}
\end{align}
Figure~\ref{fig:pp4} illustrates the evolution of the spin frequency due to superradiance for different axion masses ($m_{A}$) and magnetic field strengths ($B$), taking the pulsar PSR J1748$-$2446ad as the reference source with an initial spin frequency of $f_{0}=716$ Hz and this analysis assumes $g_{A\gamma\gamma}=10^{-12}$ GeV$^{-1}$ and $\sigma_{M}R = 0.1$~\cite{Hessels:2006ze, Day:2019bbh}. The results indicate that larger axion masses and stronger magnetic fields (shown through the colour bar) lead to a more rapid spin-down of the pulsar due to enhanced superradiance effects.
%

%
\subsection{Combined effect and braking index evolution}
Considering that a neutron star can undergo multiple spin-down mechanisms simultaneously, the total torque acting on the star can be given as,
\begin{align}
	I\dot{\Omega}_{m}=N_{\text{total}}
	=N_{\text{EM}} + N_{\text{GW}} + N_{r} + N_{\text{SUP}},
\end{align}
where, from Eqs.~(\ref{2.1}), (\ref{2.3}), (\ref{2.8}) and (\ref{2.17}), we obtain the individual contributions from magnetic dipole radiation ($N_{\text{EM}}$), gravitational wave emission ($N_{\text{GW}}$), resistive processes ($N_{r}$), and superradiance ($N_{\text{SUP}}$), respectively. The quantity $\dot{\Omega}_{m}$ denotes the total spin-down rate arising from all mechanisms.
\begin{figure}[t]
	\centering
	\includegraphics[scale=0.5]{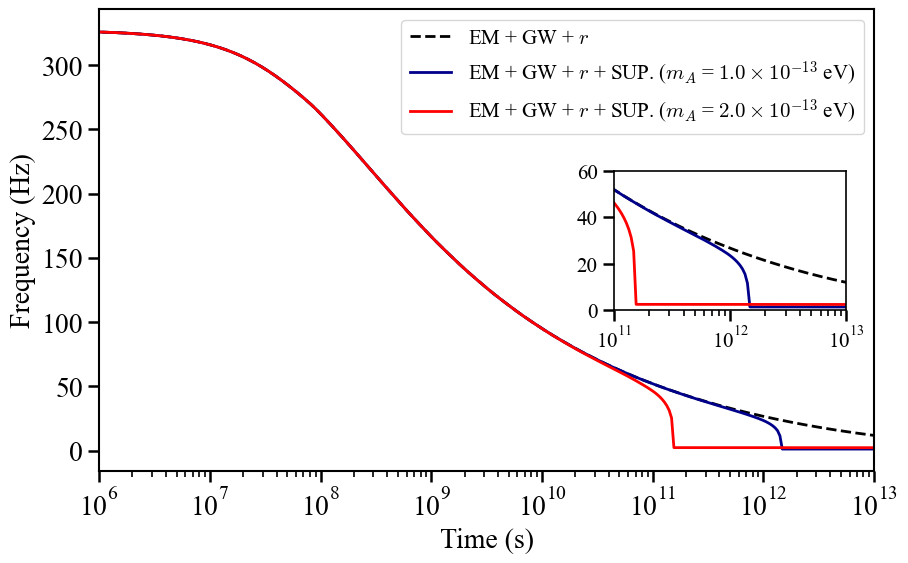}
	\caption{Comparative analysis of multi-mechanism spin-down, including superradiance and conventional torque processes, assuming a magnetic field $B=10^{13}\,\mathrm{G}$, $r$-mode amplitude $\alpha=10^{-4}$, and ellipticity $\epsilon \approx 10^{-6}$~\cite{Andersson:2000mf}.}
	\label{fig:pp5}
\end{figure}
Figure~\ref{fig:pp5} compares the spin evolution including superradiance with the conventional torque contributions, taking as reference the isolated pulsar PSR J1824$-$2452A~\cite{Cipolletta:2015nga}. The inclusion of superradiance leads to a significantly enhanced spin-down rate, resulting in a rapid decline of the stellar rotation frequency. Once the superradiant phase ceases, the evolution transitions smoothly back to the standard spin-down regime dominated by electromagnetic and gravitational torques.
We systematically quantify the spin-down evolution by computing the time-dependent braking index defined as
\begin{align}
	n(t) = \frac{\Omega(t)\,\ddot{\Omega}(t)}{\dot{\Omega}(t)^2}.
\end{align}
The evolution of the braking index as a function of time is shown in Fig.~\ref{fig:braking}, where we compare the standard multi-component spin-down model (EM+GW+$r$) with the case including superradiance (EM+GW+$r$+SUP).
In the combined evolution (EM+GW+$r$+SUP), the superradiant contribution is present throughout the early and intermediate stages of the spin-down, as long as the rotational frequency satisfies $\Omega > \omega$, where $\omega$ denotes the critical frequency below which the superradiant channel becomes inactive. Consequently, the observed features in the braking index arise from the continuous interplay between superradiance and the standard spin-down mechanisms.
At early times, the neutron star rotates rapidly and all torque components, including superradiance, contribute simultaneously to the spin evolution. In this regime, the electromagnetic, gravitational-wave, and $r$-mode torques exhibit a strong dependence on the rotational frequency, resulting in an initially efficient spin-down. However, the superradiant contribution possesses a qualitatively different frequency dependence, which effectively reduces the sensitivity of the total torque to variations in the rotation rate. As a result, although the star continues to spin down, the acceleration of the spin-down decreases, corresponding to a phase in which the magnitude of the spin-down rate weakens. This behavior is reflected in the braking index, which decreases and can attain negative values.
\begin{figure}[t]
	\centering
	\includegraphics[scale=0.45]{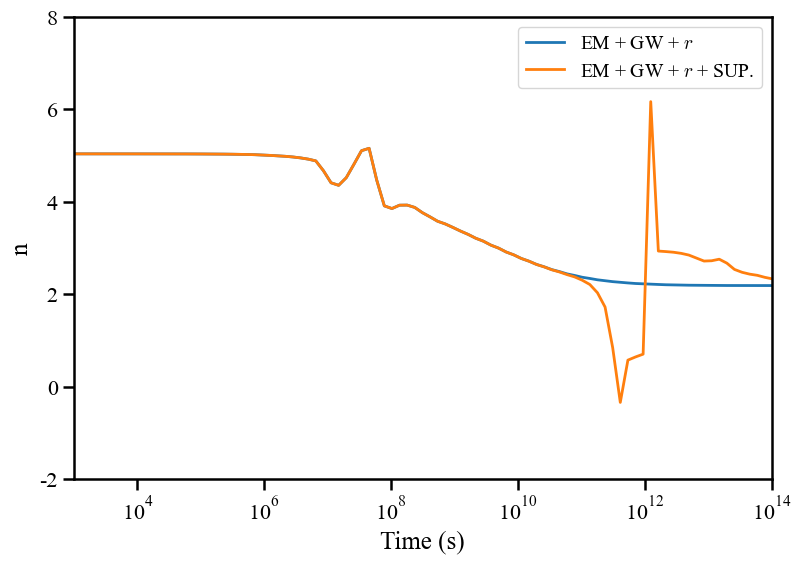}
	\caption{Evolution of the braking index $n(t)$ with and without the inclusion of superradiance, considering the same parameters as in Figure~\ref{fig:pp5}.}
	\label{fig:braking}
\end{figure}
As the evolution proceeds and the rotational frequency decreases, the relative contributions of the different torque components evolve. While the higher-order standard terms diminish more rapidly, the superradiant contribution remains comparatively significant, leading to a regime of strong competition between the various spin-down channels. In this intermediate phase, the structure of the total torque changes such that its dependence on frequency becomes steeper. This enhances the efficiency of angular momentum extraction and results in a rapid increase of the braking index, producing large positive values.
Finally, when the rotational frequency drops below the critical value $\omega$, the superradiant channel becomes inactive. Beyond this point, the evolution is governed solely by the standard electromagnetic, gravitational-wave, and $r$-mode torques. Consequently, the braking index relaxes and asymptotically approaches the value determined by these conventional spin-down mechanisms.


\section{Gravitational wave spectra}
\label{sec:gwSpec}
Gravitational waves (GWs) from an isolated neutron star can arise through several mechanisms. Two of the most prominent are: (i) direct GW emission produced by non-axisymmetric deformations of the stellar structure, and (ii) radiation generated by the excitation of $r$-mode oscillations. In addition, axion clouds formed through superradiant instability can also serve as a continuous source of gravitational waves. In this section, we discuss these various channels of GW emission from an isolated neutron star.
\subsection{GW signal from quadrupolar deformation and fluid oscillation}
Considering the angular momentum loss due to the $r$-mode instability~\cite{Shapiro:1983du}, the dominant multipole contribution arises from the mode with $l=m=2$. For this channel the time averaged strain amplitude at a distance $D$ from the source can be expressed as~\cite{Owen:1998xg},
\begin{equation}
	h_{r}(t) = \frac{256}{45}\sqrt{\frac{\pi}{30}} \frac{GMR^{3}}{Dc^{5}} \, \Tilde{J} \, \alpha \, \Omega^{3},
\end{equation}
where $M$ and $R$ are mass and radius of the star, $\alpha$ is the dimensionless $r$-mode amplitude, and $\Tilde{J}$ is the dimensionless constant.  
The corresponding characteristic strain amplitude of the $r$-mode GWs signal~\cite{Ho:1999fh} can be written as,
\begin{equation}
	h_{c(r)} = f_{r} \, h_{r} \, \sqrt{\left| \frac{dt}{df_{r}} \right|},
\end{equation}
with the gravitational-wave frequency associated with the $r$-mode $f_{r} = 2\Omega/3\pi$.
In a parallel way, GWs may also arise from a time varying quadrupolar deformation of a rotating neutron star. For a triaxially deformed star, the strain amplitude at a distance $D$ takes the form~\cite{Huang:2022vnz},
\begin{equation}
	h_{g}(t) = \frac{4 G I \epsilon}{D c^{4}} \, \Omega^{2},
\end{equation}
where $I$ represents the star's moment of inertia and $\epsilon$ measures the equatorial ellipticity characterizing the deviation from perfect axi-symmetry. 
\begin{figure}[t]
	\centering
	\includegraphics[scale=0.60]{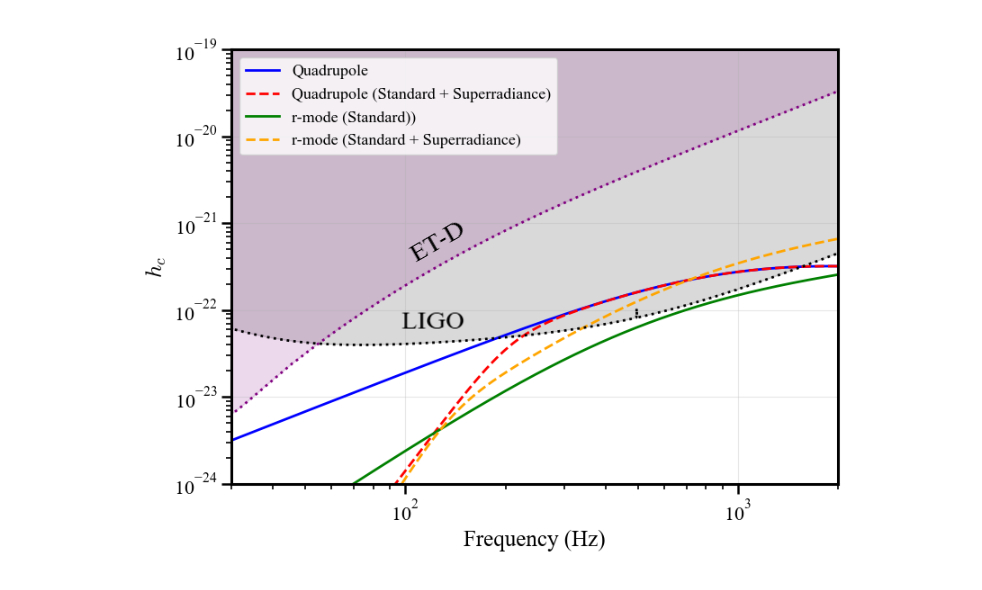}
	\caption{Characteristic strain as a function of frequency under different astrophysical conditions.}
	\label{fig:pp6}
\end{figure}
The associated GW frequency for this mechanism is related to the rotational frequency by $f_{g} = \Omega/\pi$ and the characteristic strain amplitude of this deformation powered emission is thus,
\begin{align}
	h_{c(g)} = f_{g} \, h_{g} \, \sqrt{\left| \frac{dt}{df_{g}} \right|}.
\end{align}
Fig.~\ref{fig:pp6} illustrates the characteristic gravitational wave strain $h_{c(r)}$ and $h_{c(g)}$ from two primary emission mechanisms in an isolated neutron star: $r$-mode oscillations and crustal deformations. The $r$-mode induced strain amplitude $h_{c(r)}$ is depicted by the yellow line, assuming $r$-mode amplitude $\alpha = 10^{-4}$. Gravitational wave amplitude strain $h_{c(g)}$ due to the star's quadrupolar deformation is shown for ellipticity value $\epsilon = 10^{-6}$ (blue solid line), with the spin-down rate derived by considering different spindown mechanisms assuming $B =10^{13}$ G (mentioned as standard spin-down process) with and without superradiance. For all cases, both $\alpha$ and $\epsilon$ are assumed to be constant over time. The black dotted line and the purple dotted line represents the Advanced LIGO~\cite{Borhanian:2022czq,Hild:2011np} and ET sensitivity curve~\cite{ET:2019dnz,Hild:2008ng} respectively, providing a reference for the detectability of such signals.

\subsection{GW frequency shift due to superradiance} 
As mentioned earlier, ultralight bosons, such as axions or axion-like particles, can accumulate around rapidly rotating neutron stars, forming massive clouds through the process of superradiant instability. When the neutron star's size is comparable to the axion's Compton wavelength, this accumulation occurs efficiently~\cite{Arvanitaki:2009fg,Arvanitaki:2010sy}. These axionic clouds, with their large occupation numbers, can act as a source of continuous gravitational wave (GW) signals. One key mechanism behind this emission is the annihilation of axions, which generates a coherent and monochromatic GW signal that persists over long durations. Prior studies have considered axion production via non-stationary pair-plasma discharges near pulsar polar caps \cite{Noordhuis:2023wid}. These axions can form gravitationally bound state around neutron stars, but their large masses are not favourable for creation of superradiant instability around neutron stars.
\begin{figure*}[h]
	\centering
	\begin{minipage}[b]{0.45\textwidth}
		\centering
		\includegraphics[width=\linewidth]{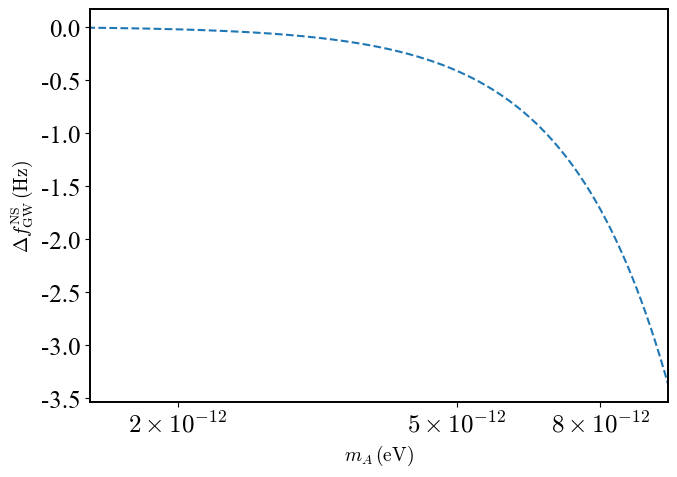}
	\end{minipage}
	\hfill
	\begin{minipage}[b]{0.50\textwidth}
		\centering
		\includegraphics[width=\linewidth]{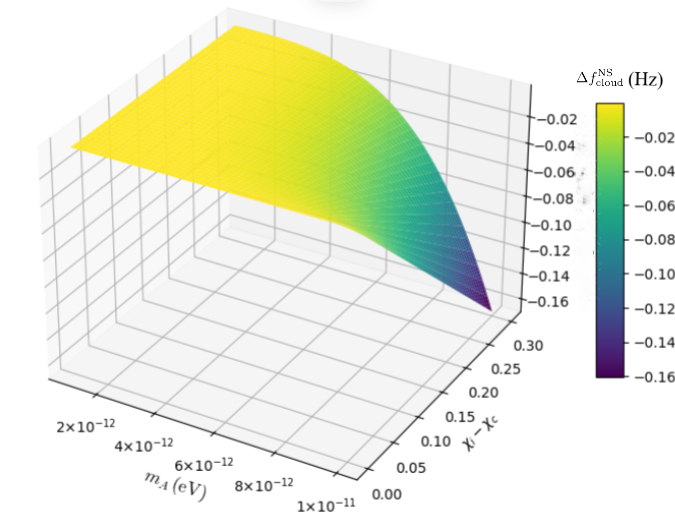}
	\end{minipage}
	\caption{(Left) Leading-order correction term to the gravitational-wave frequency arising from the presence of axion clouds. (Right) Contribution of the cloud itself to the frequency correction for different neutron star masses and spin parameters.}
	\label{fig:pp10}
\end{figure*}
%
At leading order, the frequency of the emitted GW signal $f^0_{\mathrm{GW}}$~\cite{Brito:2014wla} is directly tied to the axion's rest energy and is given by:
\begin{equation}f^0_{\mathrm{GW}}=\frac{2m_{A}}{h}\approx48.3\,\mathrm{Hz}\left(\frac{m_{A}}{10^{-13}\,\mathrm{eV}}\right),
	\vspace{0.3cm}
\end{equation}
where $m_{A}$ is the axion mass. In this study, we specifically focus on the gravitational wave signals produced by axions  (considering annihilation signals only from the fastest-growing bound state) interacting gravitationally with neutron stars. These signals, if detected, could offer a fascinating window into the existence and properties of axions, potentially shedding light on new physics beyond the Standard Model. For axions, the growth is exponential and results in a cloud with a macroscopic number of particles all occupying the same state. There are corrections to the signal frequency due to gravitional potential of the neutron star and the cloud itself. The frequency shift due to gravitational potential $\Delta{f^{\mathrm{NS}}_{\mathrm{GW}}}$ and the frequency shift due to cloud $\Delta{f^{\mathrm{NS}}_{\mathrm{cloud}}}$ are defined as~\cite{Zhu:2020tht,Baumann:2018vus},
\begin{gather}
	\Delta{f^{\mathrm{NS}}_{\mathrm{GW}}}=m_{A}\left(\frac{\alpha_{0}^2}{2(l^{\prime}+n^{\prime}+1)^2}+\frac{\alpha_{0}^4}{8(l^{\prime}+n^{\prime}+1)^4}\right),\\
	\Delta{f^{\mathrm{NS}}_{\mathrm{cloud}}}=0.2f^{0}_{\mathrm{GW}}\alpha_{0}^{2}\frac{M_{\mathrm{cloud}}}{M_{\mathrm{NS}}},\\
	f_{\mathrm{GW}}=f^{0}_{\mathrm{GW}}-\Delta{f^{\mathrm{NS}}_{\mathrm{GW}}}-\Delta{f^{\mathrm{NS}}_{\mathrm{cloud}}},
\end{gather}
where $M_{\mathrm{NS}}$ and $M_{\mathrm{cloud}}$ denote the masses of neutron star and cloud, respectively. The left panel of Fig.~\ref{fig:pp10} shows how the frequency shift due to gravitational potential $\Delta{f^{\mathrm{NS}}_{\mathrm{GW}}}$changes with axion mass $m_{A}$. The right panel of Fig.~\ref{fig:pp10} shows the three dimensional visualization of frequency shift due to cloud for different axion masses and spin parameters $\chi_{i} -\chi_{c}$, where the colour bar also represents the frequency shift.

%
\subsection{GW strain from cloud around neutron stars}
Among the possible bound states, the fastest growing superradiant mode corresponds to the first unstable level which is characterised by,
\begin{equation}
	(n^{\prime},l^{\prime},m^{\prime})=(0,1,1).
\end{equation}
This mode extracts rotational energy from the neutron star most efficiently. The rotational state of the neutron star can be characterized by the dimensionless spin parameter,
\begin{equation}
	\chi_{i}(t)=\frac{J(t)}{G M^{2}}
	=\frac{I\Omega(t)}{G M^{2}},
\end{equation}
where $J(t)$ is the angular momentum of the neutron star, $I$ is its moment of inertia, and $\Omega(t)$ is the angular frequency. The threshold for the instability is determined by the critical spin parameter $\chi_{c}$, given approximately by~\cite{Day:2019bbh},
\begin{equation}
	\chi_{c}\approx m_{A}
	\left(
	1-\frac{\alpha_{0}^{2}}{8}
	\right).
\end{equation}
A cloud of axions with rest energy $m_{A}$ will form around a neutron star of mass $M_\mathrm{{NS}}$ if $\chi_{i} > \chi_{c}$, where $\chi_{i}$ actually depends on the rotational spin frequency of the neutron star. For superradiantly unstable state, the instability will cause the neutron star to transfer mass and spin to the scalar field until the system reaches the saturation point. Neglecting accretion for simplicity, the evolution of the system is governed by~\cite{Brito:2017zvb},
\begin{equation}
	\dot{M}_{\mathrm{NS}} = -\dot{E}_{\mathrm{cloud}},
\end{equation}
\begin{equation}
	\dot{M}_{\mathrm{NS}} + \dot{M}_{\mathrm{cloud}} = -\dot{E},
\end{equation}
where $E$ and $E_{\mathrm{cloud}}$ are the total rotational energy of the star and the cloud respectively. The gravitational waves are generated approximately in a background determined by the final mass of the star. As the axions annihilate to gravitons and the cloud gets depleted, the GW signal amplitude $h_{0}$ gradually decreases from its maximum value. During this process, the cloud extracts angular momentum from the star until the stellar spin parameter satisfies $\chi_{i} \approx \chi_{c}$, at which point the growth of the cloud effectively halts. The final masses of the neutron star and the axion cloud after the cessation of superradiance can be defined as~\cite{Zhu:2020tht},
\begin{align}
	M_\mathrm{{NS,f}} &= M_\mathrm{{NS}}\left[1-\alpha_{0}(\chi_{i}-\chi_{c})\right],\\
	M_\mathrm{{cloud,f}} &= M_\mathrm{{NS}}\alpha_{0}(\chi_{i}-\chi_{c}).
\end{align}
In the small $\alpha_0$ regime ($\alpha_0 \ll 1$), the gravitational-wave power emitted by the cloud is given by~\cite{Brito:2014wla},
\begin{equation}
	P_\mathrm{{GW}} \approx  0.025 \, \frac{c^5}{G} \, \alpha_0^{14}
	\left( \frac{M_{\mathrm{cloud,f}}}{M_{\mathrm{NS,f}}} \right)^2.
	\label{eq2}
\end{equation}
The gravitational-wave strain amplitude can be obtained from the energy flux carried by the wave. At a distance $r$ from the source, the flux is related to the emitted power through,
\begin{equation}
	F = \frac{P_\mathrm{{GW}}}{4\pi r^2},
\end{equation}
while the gravitational-wave energy flux may be written as
\begin{equation}
	F = \frac{c^3}{16\pi G}
	\left\langle \dot{h}_+^2 + \dot{h}_\times^2 \right\rangle,
\end{equation}
and performing the time and angular average for a monochromatic wave, one obtains,
\begin{equation}
	h_{0,\mathrm{peak}} =
	\left(
	\frac{5 G P_\mathrm{GW}}{2c^3 f_\mathrm{{GW}}^2\pi^2 r^2}
	\right)^{1/2}.
	\label{eq1}
\end{equation}
Now using Eqs.~\eqref{eq2} and~\eqref{eq1}, and adopting a canonical neutron star with mass $M_{\rm NS}=1.4\,M_\odot$ located at a distance of $r=1\,{\rm kpc}$, the leading-order peak strain is given by,
\begin{equation}
	h_{0, \mathrm{peak}}\approx4.2\times10^{-25}\left(\frac{0.105\,m_{A}}{10^{-12}\,\mathrm{eV}}\right)^{7}\left(\frac{\chi_{i}-\chi_{c}}{0.5}\right).
\end{equation}
\begin{figure}[t]
	\centering
	\includegraphics[scale=0.5]{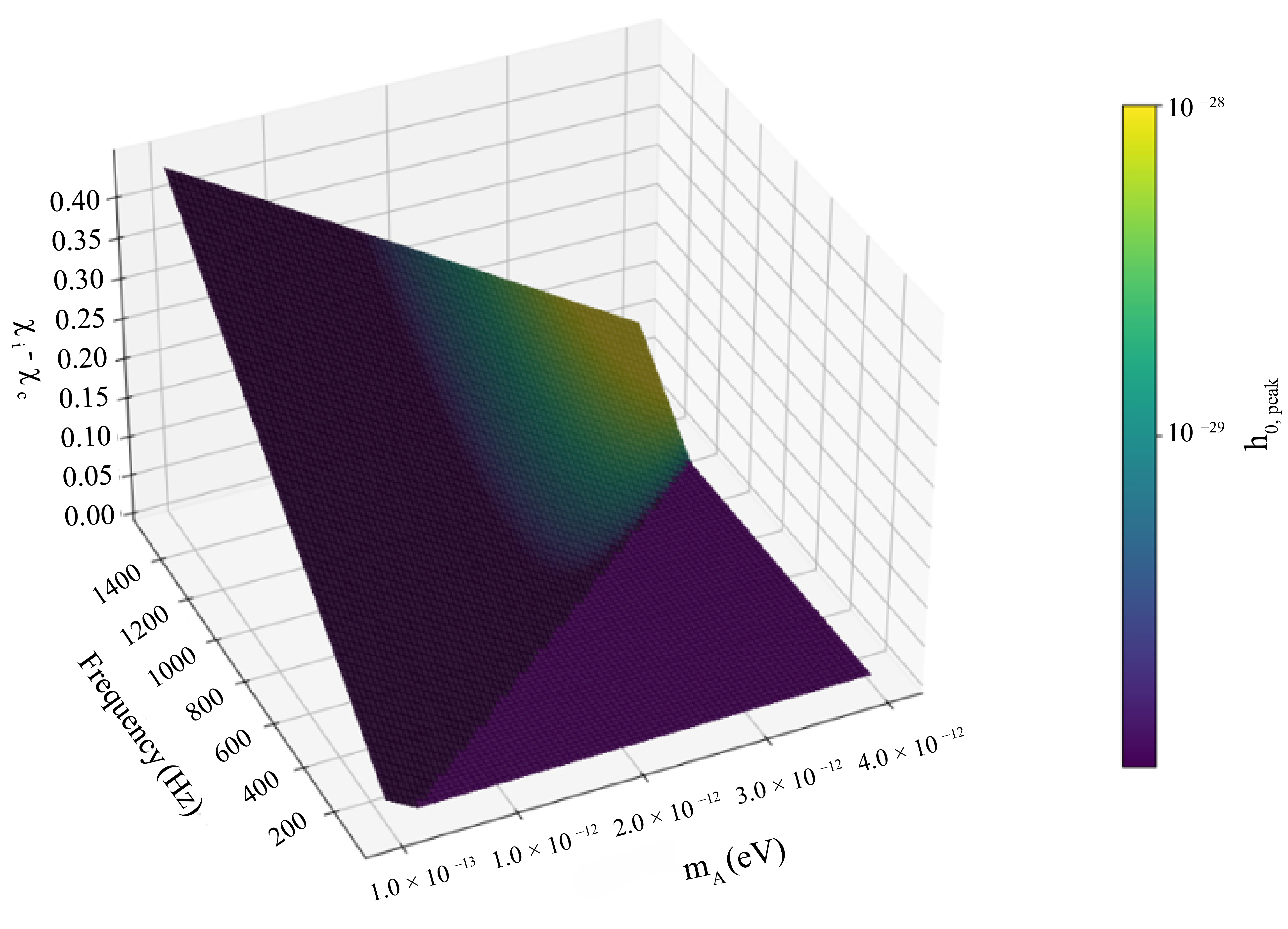}
	\caption{Characteristic strain as a function of spin-parameter and axion masses.}
	\label{fig:pp7}
\end{figure}
Fig.~\ref{fig:pp7} presents a three-dimensional visualization of how the star's spin frequency and the corresponding change in spin parameter ($\chi_{i} -\chi_{c}$) vary with axion mass. This plot shows that across the axion mass range considered, and for spin frequencies up to 1400 Hz, the superradiance-driven axion clouds around a neutron star produce a peak gravitational wave strain amplitude are in the order $h_{0, \mathrm{peak}}\sim 10^{-28}$. The colour bar represents the peak gravitational wave strain amplitude produced by superradiance induced axion clouds, highlighting its dependence on all relevant parameters.


\section{Summary and Conclusion}
\label{sec:concl}
The spin of neutron star is an observable quantity that provides valuable insight into various physical phenomena occurring around it. Standard spin-down mechanisms -- such as electromagnetic radiation, gravitational-wave emission, and $r$-mode oscillations -- correspond to canonical braking index values of 3, 5, and 7, respectively. In reality, the spin evolution of neutron stars are far more complex. Some stars also have transient spin-up episodes, which may arise from fallback accretions. In addition to these conventional spin-down mechanisms, a rare class of events -- enhanced spin-down rate has also been observed. In this work, we have explained superradiant instability as an efficient spin-down process, whose effectiveness is governed by axion mass and by the bulk electrical conductivity of the neutron star in the presence of a strong magnetic field and also it can express higher and negative braking index as observed in several pulsars. As the superradiant phase operates over a timescale much shorter than star's lifetime, the neutron star subsequently returns to its conventional spin-down evolution once the instability ceases.
An important outcome of our analysis is the non-trivial evolution of the braking index during the superradiant phase. Unlike the canonical values predicted by conventional torque mechanisms, the inclusion of superradiance leads to a strongly time-dependent braking index that can attain both significantly enhanced positive values and transient negative values. Such behaviour emerges naturally from the competition between the conventional spin-down torques and the additional angular-momentum extraction induced by superradiance. This provides a possible explanation for several pulsars and magnetars that exhibit anomalous braking-index measurements, including negative braking indices and unusually large positive values that are difficult to reconcile within standard spin-down scenarios alone.
From a gravitational-wave perspective, the coexistence of multiple emission mechanisms, most notably $r$-mode oscillations and quadrupole deformations, directly shapes the predicted characteristic strain from an isolated neutron star. In a multi-messenger context, these mechanisms offer complementary probes of the star’s internal physics. Although these signals arise from fundamentally different physical processes, they can operate simultaneously and independently at any given moment, generating a multi-frequency GW spectrum from the same object, which are within the reach of current ground based detectors like LIGO and ET. In the context of GW due to superradiance, our analysis indicates that the peak gravitational-wave strain from axion clouds associated with superradiant spin-down remains at the level of $h_0 \sim 10^{-28}$, placing it beyond the sensitivity of current detectors but within the reach of next-generation gravitational-wave observatories.


%
 
\acknowledgments
UKD acknowledges support from the Anusandhan National Research Foundation (ANRF), Government of India under Grant Reference No.~CRG/2023/003769.

\bibliographystyle{JHEP}
\bibliography{nsspindwnRef.bib}

\end{document}